\documentclass[aps,twocolumn,showpacs,showkeys,floatfix]{revtex4}

\usepackage{amsfonts}
\usepackage{amsmath}
\usepackage{amssymb}
\usepackage{graphicx} 
\usepackage{xcolor}
\usepackage{dcolumn}
\usepackage{bm}
\usepackage{hyperref}
\begin{document}

\preprint{APS/123-QED}

\title{Radio-frequency induced Autler-Townes Effect for single- and double-photon magnetic-dipole transitions in the Cesium ground state}

\author{Arturs Mozers}%
\email{arturs.mozers@lu.lv}
\author{Linda Serzane-Sadovska}
\author{Florian Gahbauer}
\author{Marcis Auzinsh}

\affiliation{
Laser Centre, Faculty of Science and Technology, University of Latvia, Jelgavas Street 3, Riga, Latvia
}

\begin{abstract}
 We have observed the Autler-Townes effect in single- and suspected double-photon magnetic-dipole transitions in the Cesium ground-state magnetic-sublevel manifold. Experiments were performed in a Cesium vapor cell. The D$_1$ line was excited by laser radiation to create ground-state optical polarization, and transitions between the ground-state magnetic sublevels were excited by radio-frequency (RF) radiation. Two different excitation geometries were studied: in one case the electric field vector of the linearly polarized laser radiation was parallel to the static magnetic field, whereas in the other case these vectors were perpendicular. The oscillating magnetic field produced by the RF coils was in the plane perpendicular to the electric field vector of the laser radiation. The Autler-Townes effect was confirmed by its linear dependence on the RF magnetic field amplitude, which is proportional to the Rabi frequency, in the case of single-photon transitions. We also observed peaks that by their position appeared to correspond to double and even triple photon transitions, which were more pronounced when the DC magnetic field and optical electric field vectors were perpendicular. In the peak at an energy that corresponds to two photons, splitting with a quadratic dependence on the RF magnetic field amplitude could be observed. The experimental measurements are supplemented by theoretical calculations of a model $J=1 \longrightarrow J=0$ system.  
\end{abstract}
\date{\today}


\date{\today}

\maketitle

\section{Introduction}
Applying optical and radio frequency (RF) fields simultaneously to a multi-level system with Zeeman splitting in a magnetic field has long been used to study the Zeeman structure~\cite{Brossel:1952}. Optical-RF double resonance techniques continue to be relevant for magnetometry applications~\cite{Weis:2006,DiDomenico:2007,Zigdon:2010,LeGal:2021,Bertrand:2021,Zhang:2022,LeGal:2022}. RF fields are simple to produce, and the external magnetic field value can be obtained easily from the position of the optical-RF double resonance peaks in absorption or fluorescence, making them very convenient for magnetometry. Moreover, with proper design, RF coils can be added to a 2-axis optical magnetometer~\cite{LeGal:2019} to convert it into a 3-axis magnetometer. As is well known when the RF field is in resonance with a magnetic sublevel transition, the population distribution generated by the optical pumping is modified as transitions induced by the RF field take place between the magnetic sublevels. The result is a peak in the absorption or fluorescence signal, and the peak position can be directly connected to the external magnetic field strength. As usual, an optimization must be performed between the signal contrast and width, both of which are increased as the RF field Rabi frequency is increased. For particularly large RF field Rabi frequencies, these peaks are split into doublets as a result of the Autler-Townes (AT) effect~\cite{Autler:1955}. Autler and Townes used a strong RF field to excite electric-dipole transitions among the $J=1$ and $J=2$ sublevels of OCS molecules and probed the resulting energy-level splitting with a weaker microwave field that connected the $J=1$ ground state with the $J=2$ excited state. Later experiments used two optical fields, where a strong pump field induced the splitting in a transition while a weak probe field was scanned across this transition to show the characteristic double peak associated with the AT effect~\cite{Gray:1978, Delsart:1976}. The magnitude of the splitting is proportional to the square root of the pump field intensity, which is to say, proportional to the pump field Rabi frequency. The AT effect was studied in an optical-RF double resonance experiment~\cite{Bechtel:1987} by pumping the Na D$_1$ line and probing with weak RF field that induced transitions in the ground-state magnetic sublevels. Again, the magnitude of the AT splitting was found to be proportional to the square root of the pump laser intensity. 

\begin{figure}[ht]
    \centering
    \includegraphics[width=\linewidth]{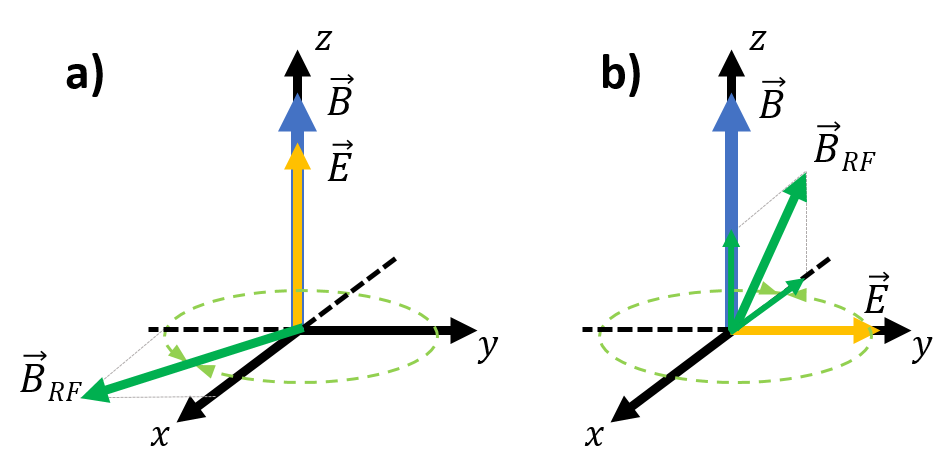}
    \caption{Excitation geometries: (a) $\mathbf{E} \parallel \mathbf{B}$; (b) $\mathbf{E} \perp \mathbf{B}$}.
    \label{fig:geometry}
\end{figure}

In this work we observed the AT splitting caused not by optical radiation that induced electric dipole transitions, but by an RF field that induced magnetic dipole transitions between the ground-state sublevels of Cesium while the D$_1$ line was pumped with  moderately intense, linearly polarized laser radiation. The transitions between the ground-state sublevels and the AT splitting of these transitions were observed by changes in the absorption of the optical light.  Two cases were studied: (i) the electric field vector $\mathbf{E}$ of the laser radiation was parallel to a DC, external magnetic field $\mathbf{B}$ ($\mathbf{E} \parallel \mathbf{B}$),  and (ii) the electric field vector of the laser radiation was perpendicular to the DC magnetic field ($\mathbf{E} \perp \mathbf{B}$). Both configurationas are shown in Fig.~\ref{fig:geometry}. In the second case, in addition to expected optical-RF double resonance transitions between ground-state magnetic sublevels, transitions were also observed at magnetic fields at which the energy difference between two adjacent magnetic sublevels corresponded to twice the RF photon energy, and these also displayed AT splitting at large amplitudes of the oscillating magnetic field $B_{RF}$. While the magnitude of the AT splitting was linear in the Rabi frequency of the RF field for the standard, single-photon transitions, it was quadratic for the transitions corresponding to twice the photon energy. The energy difference and the quadratic dependence of the AT splitting on the Rabi frequency lead us to suspect that these peaks could be attributed to double-photon transitions, although as yet we have not been able to develop a numerical model to explain this little-studied effect. 

For the standard case, the splitting between the two peaks $\Delta E$ is given by
\begin{equation}
    \Delta E = E_{+}-E_{-}=\hbar \Omega_{RF}=\mu B_{RF}^0,
\end{equation}
where $E_{+}$ and $E_{-}$ are the positions of the right and left peaks, respectively, and $\Omega_{RF}$ is the Rabi frequency associated with the amplitude of the RF field $B_{RF}^0$, and $\mu=\langle i |\hat{\mu} | j \rangle$ is the magnetic dipole transition strength between sublevels $|i \rangle$ and $|j \rangle$ in the hyperfine level under consideration.

Finally, in both cases, we also observed a peak that corresponds to a transition that skipped a magnetic sublevel, which would normally be forbidden, being nominally a $\Delta m=2$ transition. However, in the geometry of case (ii) perhaps such a transition was allowed because of the magnetic sublevel mixing caused by the RF magnetic field perpendicular to the quantization axis. In addition to the experimental measurements, we have created a theoretical model based on the Liouville equation and implemented this model in the Julia language using the \verb|QuantumOptics.jl| package~\cite{kramer:2018}. The model is implemented for a $J=1 \longrightarrow J=0$ transition in order to calculate at least the single-photon AT effect and demonstrate qualitative agreement with the experiments.

\section{Experimental}
\begin{figure}[ht]
    \centering
    \includegraphics[width=\linewidth]{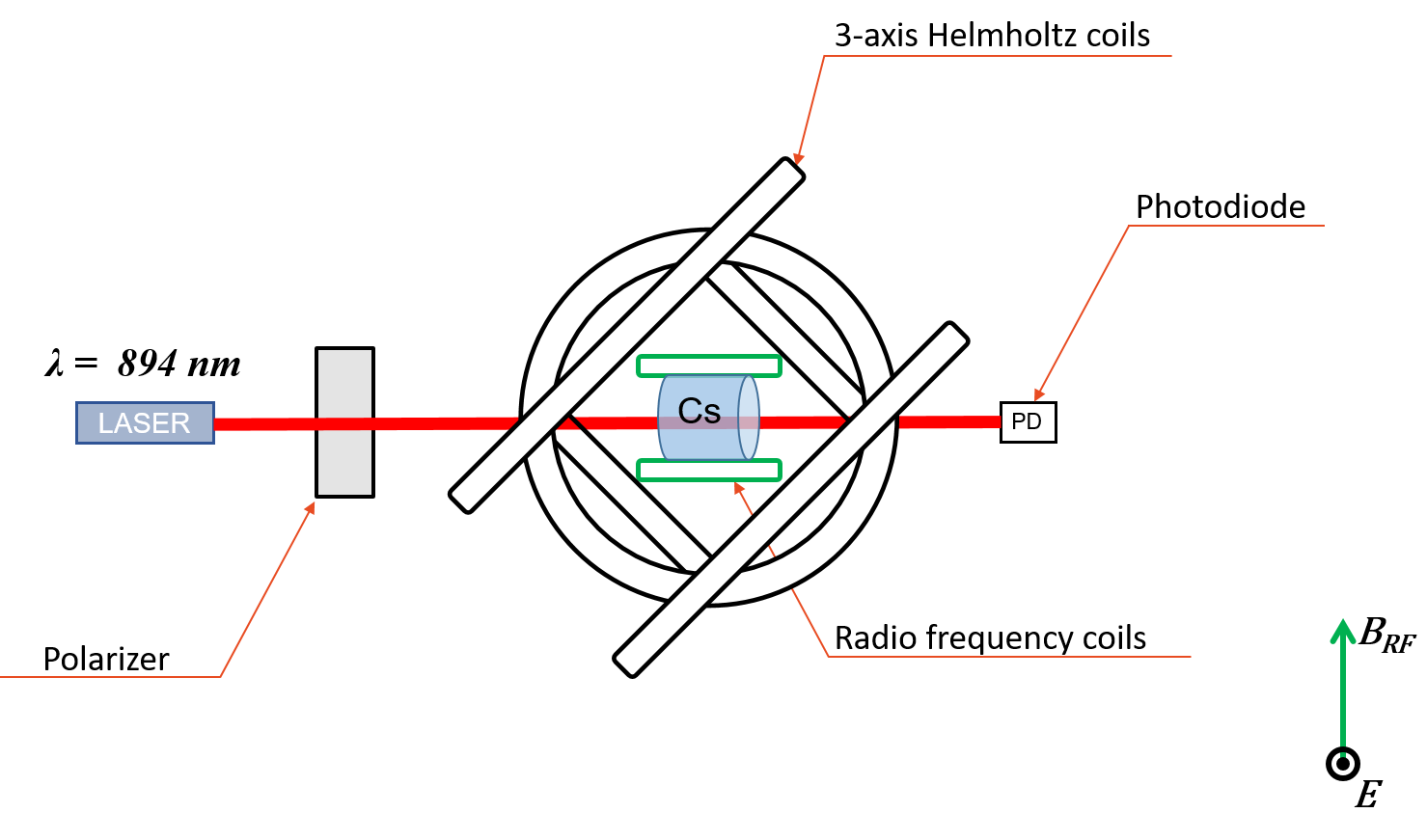}
    \caption{Experimental setup. Note the direction of the RF field and the laser polarization in the bottom right.}
    \label{fig:experiment}
\end{figure}

The experimental setup is shown in Fig.~\ref{fig:experiment}. A distributed feedback laser (Toptica DL-DFB) provided laser radiation at 50--70 $\mu$W, which passed through a linear polarizer that polarized the laser radiation along the $z$-axis. The diameter of the beam was approximately 2 mm, based on the effective diameter reported by a Coherent LASERCAM HR beam profiler. We estimate that a resonant optical field of this intensity on the Cs D$_1$ transition corresponds to a Rabi frequency of a few MHz~\cite{Mozers:2023}. A Cesium vapor cell at room temperature was placed in the center of a 3-axis Helmholtz coil system with inner diameters of 34cm, 43cm, and 52cm, respectively, for each pair of coils. These coils compensated the ambient magnetic field and scanned the DC magnetic field with a magnitude of several Gauss along the $z$-axis as indicated in Fig.~\ref{fig:geometry}. This figure also gives the geometry of the polarizations of the optical electric field vector and the RF magnetic field vector. The RF field was produced by two smaller coils in Helmholtz configuration, which surrounded the vapor cell and provide an oscillating magnetic field with the orientation shown in Fig.~\ref{fig:geometry}. The RF coils had a radius of 3~cm and were separated by 3~cm. They consisted of 30 turns of wire with a resistance of 1.21 ohm and an inductance of 0.227 mH. They were excited by a function generator connected in series through a capacitor with capacitance $C=470 \mathrm{pF}$ that can deliver a sinusoidal voltage $V(t)=V_0 \sin{\omega_{RF}t}$, where $V_0$ could be up to 5~V and $\omega_{RF}/(2\pi)$ is the oscillation frequency, which was held constant at 508~kHz in this study. Depending on the values of $\omega_{RF}$ and $C$, an oscillating magnetic field $B_{RF}(t)=B_{RF}^0 \sin{\omega_{RF}t}$ was created with $B_{RF}^0$ values up to approximately $2.3$~Gauss. The laser radiation transmission signals were obtained by slowly scanning the DC magnetic field with a frequency of 0.5~Hz. The laser beam was directed onto a photodiode after passing through the cell and the signal was recorded and averaged on an oscilloscope. 

\section{Results and Discussion}
\subsection{Excitation geometry \texorpdfstring{$\mathbf{E} \parallel \mathbf{B}$}{Lg}}
\begin{figure}[ht]
    \centering
    \includegraphics[width=\linewidth]{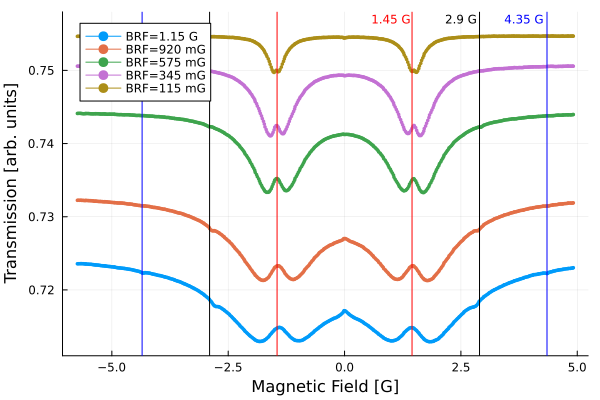}
    \caption{Experimental measurements on the Cs D$_1$ $F_g=4 \longrightarrow F_e=4$ transition for various values of the RF magnetic field amplitude. The thin vertical lines correspond to the magnetic field values shown at the top of the figure.
    }
    \label{fig:BparE-5vpp}
\end{figure}

Figure~\ref{fig:BparE-5vpp} shows plots of the experimentally observed transmission signal versus the external magnetic field for various values of the RF magnetic field amplitude. The RF magnetic field amplitude is proportional to the Rabi frequency. However, the constant of proportionality is different for the different magnetic sublevel transitions within the same manifold, and so we give only the magnetic field amplitude in the legend. Although we give the full amplitude of the oscillating magnetic field, we note that in this geometry (Fig.~\ref{fig:geometry} a), the linearly oscillating RF field is decomposed into two circularly counter-rotating components, one with photons carrying angular momentum component (projected along the $z$-axis) $q=1$ while the photons corresponding to the other component carry angular momentum component $q=-1$. The optical-RF double resonance peak was at a magnetic field value of 1.45~G, which corresponds to an RF frequency $\omega_{RF}/(2\pi)=508$~kHz. The splitting of the peaks can be seen clearly. Furthermore, for values of $B^0_{RF} \ge 920 \mathrm{mG}$, we saw an additional peak at twice the transition resonance frequency, and for $B^0_{RF} = 1.15 \mathrm{G}$ even at 3 times the resonance frequency, which we think may be attributed to 2-photon and 3-photon transitions, respectively.

\begin{figure}[ht]
    \centering

    \includegraphics[width=\linewidth]{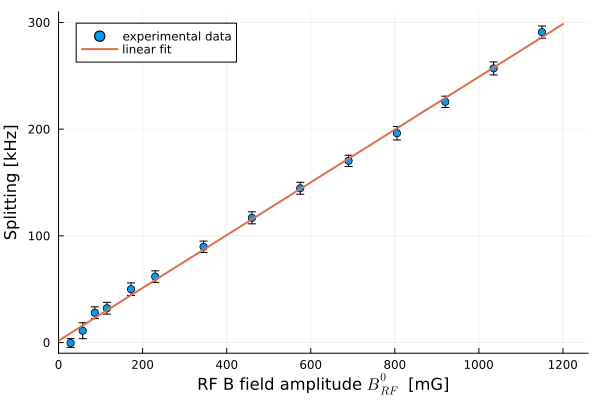}
    \caption{Autler-Townes splitting versus RF magnetic field amplitude. Cs D$_1$ $F_g=4 \longrightarrow F_e=4$ transition, $\mathbf{E} \parallel \mathbf{B}$.}
    \label{fig:BparE-AT-vs-Rabi}
\end{figure}

\begin{figure}[ht]
    \centering
    \includegraphics[width=\linewidth]{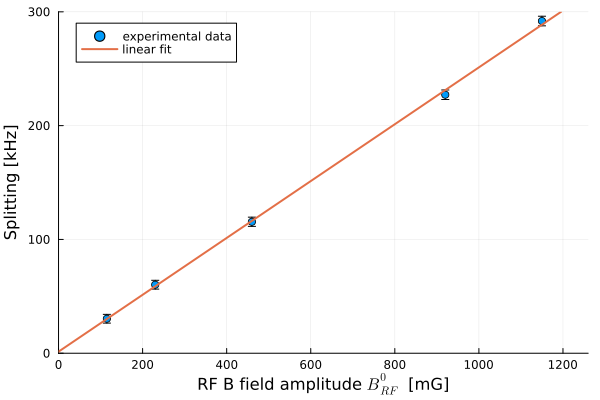}
    \caption{Autler-Townes splitting versus RF magnetic field amplitude. Cs D$_1$ $F_g=3 \longrightarrow F_e=3$ transition, $\mathbf{E} \parallel \mathbf{B}$.}
    \label{fig:BparE-AT-vs-Rabi-3-3}
\end{figure}

To test if the splitting of the peaks corresponded to AT splitting, the magnitude of the splitting is plotted against the RF magnetic field amplitude in Fig.~\ref{fig:BparE-AT-vs-Rabi} for the $F_g=4\longrightarrow F_e=4$ transition. The relationship is perfectly linear, as expected. Fig.~\ref{fig:BparE-AT-vs-Rabi-3-3} shows the same plot for the $F_g=3\longrightarrow F_e=3$ transition.  We note that the values of the RF magnetic field amplitude may contain a non-negligible systematic error because of inaccuracies in the measurement of the capacitance and inductance of the RF coils.  Still, the two-parameter linear fit approximates the data well and passes through the origin to within experimental uncertainty, as expected. The results for the $F_g=4 \longrightarrow F_e=3$ transition were similar to the $F_g=4 \longrightarrow F_e=4$ transition. The signals for the $F_g=3 \longrightarrow F_e=4$ transition were small when compared to signals for other Cs D$_1$ transtions and difficult to interpret. 

\subsection{Excitation geometry \texorpdfstring{$\mathbf{E} \perp \mathbf{B}$}{Lg}}

\begin{figure*}[ht]
    \centering
    \includegraphics[width=0.45\linewidth]{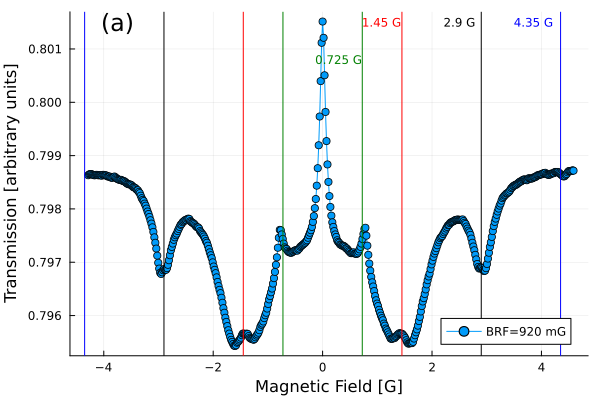}
    \includegraphics[width=0.45\linewidth]{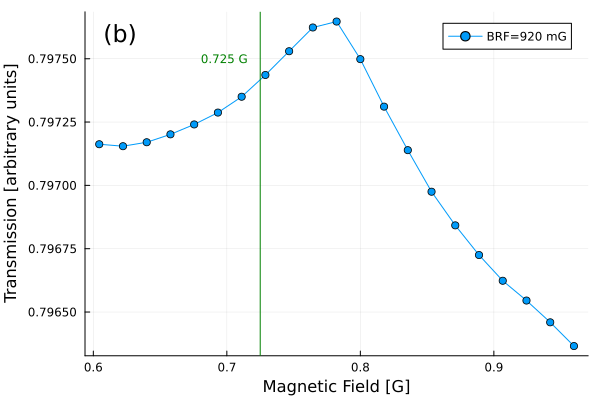}
    \includegraphics[width=0.45\linewidth]{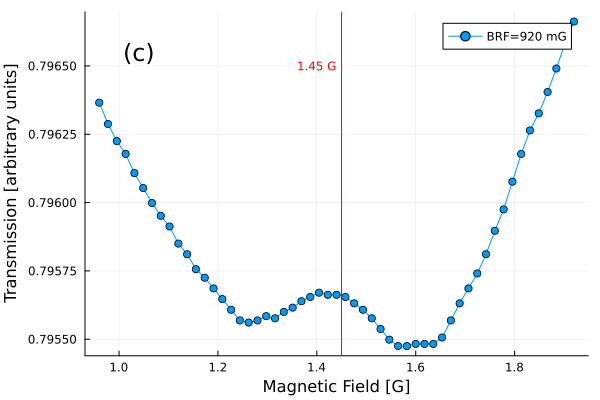}
    \includegraphics[width=0.45\linewidth]{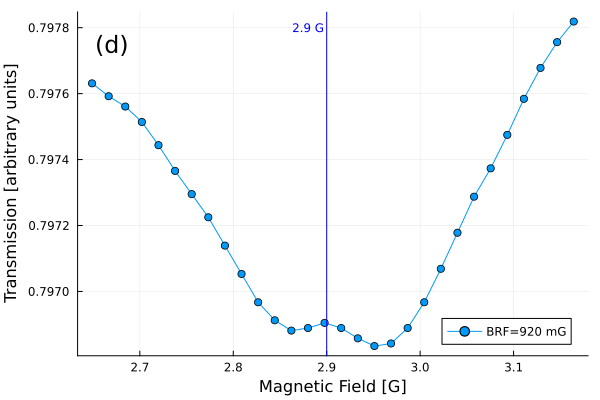}
    \caption{Cs D$_1$ $F_g=4 \longrightarrow F_e=4$ transition. $\mathbf{E} \perp \mathbf{B}$. The RF magnetic field amplitude was $B^0_{RF}=920\,\,\mathrm{mG}$. (a) Full scan from -4~G to +4~G; (b) peak around $B=0.725$~G; (c) single-photon peak around at 1.45~G; (d) peak at B= 2.9~G.} 
    \label{fig:BperpE-8vpp}
\end{figure*}

Figure~\ref{fig:BperpE-8vpp} shows the transmission signal as a function of magnetic field for the $\mathbf{E} \perp \mathbf{B}$ configuration with an estimated RF magnetic field amplitude of $B^0_{RF}=920\,\mathrm{mG}$. Fig.~\ref{fig:BperpE-8vpp}(a) shows the entire scan. Multiple peaks are visible. At a value of zero magnetic field, a Hanle resonance was observed as expected for this geometry~\cite{Hanle:1924}. Given that the RF frequency was $\omega_{RF}/(2\pi)=508$ KHz, the optical-RF single-photon peak is expected at 1.45~G, and, indeed, in the zoomed in image in Fig.~\ref{fig:BperpE-8vpp}(c) such a peak is seen with apparent AT splitting, though centered closer to 1.39~G. Additional peaks are visible near 0.725~G  and 2.9~G, as can be seen in Figs.~\ref{fig:BperpE-8vpp}(b) and (d), respectively. The latter would seem to correspond to a 2-photon transition, whereas the former seems to correspond to a transition between non-adjacent magnetic sublevels. We note that the splitting in Figs.~\ref{fig:BperpE-8vpp}(c) and (d) are asymmetrical. This asymmetry was observed over a wide range of RF field Rabi frequencies and also in some theoretical calculations using a preliminary model of an $F=4 \longrightarrow F=4$ transition. The cause of this asymmetry warrants further investigation, for which we plan to develop the theoretical model further. 

We note that the possible observation of two-photon transitions was a fortuitous result of the particular excitation geometry as shown in Fig.~\ref{fig:geometry}(b). As can be seen $\mathbf{B}_{RF}$ is neither parallel nor perpendicular to the quantization axis (the $z$-axis). As a result, the oscillating field $\mathbf{B}_{RF}$ creates photons that carry angular momentum component $q=0$ in the case of the magnetic field component that oscillates parallel to the quantization axis and angular momentum component $q=\pm 1$ for photons produced by the component of $\mathbf{B}_{RF}$ that oscillates in the plane perpendicular to the quantization axis. In the latter case they can be decomposed into two counter-rotating fields. In this way, a $q=0$ and a $q=\pm 1$ photon could be combined to produce a transition between magnetic sublevels with $\Delta m=\pm 1$. Had the RF magnetic field been oscillating perpendicularly $\mathbf{B}$, \textit{i.e.}, along the $x$-axis, then all RF photons would have had angular momentum $q=\pm 1$, and there would be no combination of two photons that could produce a $\Delta m= \pm 1$ transition. 

Another surprising peak was observed at a DC magnetic field value of $\mathrm{B}=0.725 \,\mathrm{G}$ in Fig.~\ref{fig:BperpE-8vpp}(b). This peak would correspond to a $\Delta m=2$ transition, which would be forbidden at these low magnetic fields, where level mixing caused by the interaction of the DC magnetic field and the hyperfine structure is very small\cite{Alnis:2000}. However, the components of the oscillating field $\mathrm{B}_{RF}$ perpendicular to the quantization axis introduce level mixing. The fact that the RF field can induce AT splitting already shows that it is strong enough to produce significant level mixing, which introduces components into the wavefunctions of the magnetic sublevels that would allow a $\Delta m=1$ transition to "skip over" a magnetic sublevel, as it were. Further investigation is needed to understand this effect quantitatively.

\begin{figure}[ht]
    \centering
    \includegraphics[width=\linewidth]{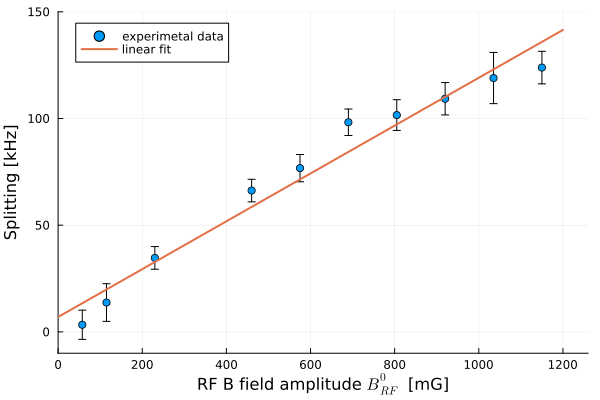}
    \caption{Autler-Townes splitting versus the RF magnetic field amplitude for the single-photon peak. Cs D$_1$ $F_g=4 \longrightarrow F_e=4$ transition, $\mathbf{E} \perp \mathbf{B}$.}
    \label{fig:BperpE-1ph-AT-vs-Rabi}
\end{figure}

The magnitude of the splitting at 1.45~G as a function of the RF magnetic field amplitude is shown in Fig.~\ref{fig:BperpE-1ph-AT-vs-Rabi}. We give the full amplitude, noting that to calculate the Rabi frequency, one would have to take into account the projection of the RF magnetic field $B_{RF}^0$ on the quantization axis (the $z$-axis). Since there was a large uncertainty in this number, we did not make this calculation, though we estimate that the angle was on the order of $45^{\circ}$.  The data were fit to a two-parameter linear function, confirming a linear relationship as in Figs.~\ref{fig:BparE-AT-vs-Rabi} and~\ref{fig:BparE-AT-vs-Rabi-3-3}.

\begin{figure}[ht]
    \centering
    \includegraphics[width=\linewidth]{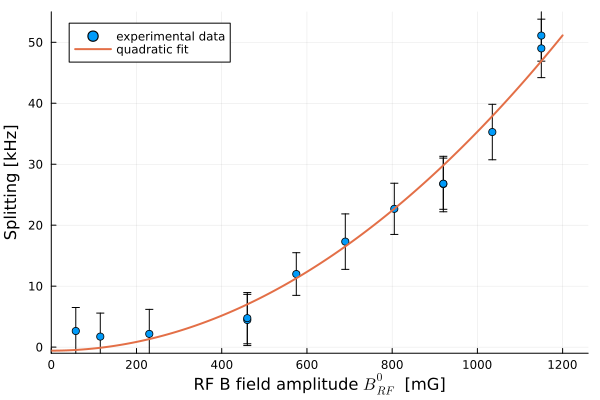}
    \caption{Autler-Townes splitting versus  the RF magnetic field amplitude for the peak at B=2.9 G. Cs D$_1$ $F_g=4 \longrightarrow F_e=4$ transition, $\mathbf{E} \perp \mathbf{B}$.}
    \label{fig:BperpE-2ph-AT-vs-Rabi}
\end{figure}

For the peak at B=2.9 G, the magnitude of the splitting as a function of RF magnetic field amplitude is shown in Fig.~\ref{fig:BperpE-2ph-AT-vs-Rabi}.  The data were well approximated with a two-parameter quadratic function $y=kx^2+b$ over the entire range. To within experimental error, the $y$-interecept was the origin. The magnetic field value at which the resonance appears and the quadratic dependence on the RF magnetic field amplitude and, thus, the Rabi frequency, suggests a two-photon process. 

\subsection{Calculations}
We implemented a theoretical model based on the work of~\cite{Zigdon:2010}, which uses a generic $J=1 \longrightarrow J=0$ transition with the excited state $|J=0, m=0 \rangle$ denoted as $e$ and the ground state sublevels $g_i=| J=1, m_i \rangle $, which correspond to $m_i=-1,0,1$ (see also~\cite{Bester:2023} for the methodology), and these are the basis states of the system. Although the experimentally studied transition was $F_g=4 \longrightarrow F_e=4$, the $J=1 \longrightarrow J=0$ model should be able to demonstrate the qualitative features.  The model is implemented as a \verb|Jupyter| notebook~\cite{Kluyver2016jupyter} using the \verb|QuantumOptics.jl| package~\cite{kramer:2018} of the \verb|Julia| programming language, which can solve the Liouville equation under steady state conditions.

In the case of the $\mathbf{E} \parallel \mathbf{B}$ geometry, the optical transitions are $\Delta m$ = 0 transitions, whereas the RF transitions are $\Delta m=\pm 1$ transitions. 
Accordingly, the interaction Hamiltonian with the electromagnetic field is given by:

\begin{align}
   \hat{H} &= \sum_{i}\Delta^{RF}_i| i \rangle   \langle i | \nonumber \\ 
& + \Omega_R \left( | 1,0 \rangle   \langle 0,0 | + | 0,0 \rangle   \left< 1,0 \right|\right ) \nonumber \\
& + \Omega_{RF} \left( | 1,1 \rangle  \langle 1,0 | + | 1,0 \rangle   \langle 1,1 | \right) \nonumber \\
& + \Omega_{RF} \left( | 1,-1 \rangle  \langle 1,0 | + | 1,0 \rangle   \langle 1,-1 | \right) ,
\end{align}
where $\Omega_R$ is the optical Rabi frequency, $\Omega_{RF}$ is the RF field Rabi frequency and $\Delta^{RF}_i$ is the detuning of the RF field transition frequency.

Spontaneous relaxation is defined using the jump operator as follows:
\begin{equation}
\hat{J}_{\Gamma}[
\hat{\rho}]=\sum_i \frac{\Gamma}{2} \left ( 2 \hat{\sigma}_{g_i} \hat{\rho} \hat{\sigma}_{g_i} - \hat{\sigma}_{g_i}^{\dagger} \hat{\sigma}_{g_i} \hat{\rho} - \hat{\rho} \hat{\sigma}_{g_i}^{\dagger} \hat{\sigma}_{g_i} \right ),
\end{equation}
where $\Gamma$ is the excited-state spontaneous decay rate, $\hat{\rho}$ is the density matrix and $\hat{\sigma}_{g_i}=| g_i \rangle \langle e |$ is the transition matrix for the decay from the $|0 0 \rangle$ excited state $e$ to the $i$-th ground-state sublevel $g_i$. 

In addition, transit relaxation due to atoms flying out of the laser beam is defined in terms of a Jump operator as:
\begin{equation}
\hat{J}_{\gamma}[\rho]=\sum_i \frac{\gamma}{2} \left ( \hat{P}_{g_i} \hat{\rho} \hat{P}_{g_i} - \hat{P}_{g_i}^{\dagger} \hat{P}_{g_i} \hat{\rho} - \hat{\rho} \hat{P}_{g_i}^{\dagger} \hat{P}_{g_i} \right ),
\end{equation}
where $\hat{P}_{g_i,g_i}= |g_i \rangle \langle g_i|$ is the projection operator for the ground-state level $| g_i \rangle$ and $\gamma$ is the transit relaxation rate. 

Then the Liouvillian matrix $L$ is formed from the Hamiltonian $\hat{H}$ and the Lindblad superoperator $\hat{L}=\hat{J}_{\gamma}+\hat{J}_{\Gamma}$ and solved for steady state conditions while taking into account transit relaxation by first subtracting the transit relaxation rate from the diagonal elements of the Liouville matrix  $L_{4(i-1)+i,4(i-1)+i}$ for $i=1,2,3,4$ and then solving the equation

\begin{equation}
L\rho=y,
\end{equation}
where $y_{i,i}=-\delta_{i,i} \lambda$ for $i=1,2,3$, and $\lambda=\gamma$ is the replenishment rate that corresponds to atoms flying into the beam from the surroundings. The steady-state density matrix is then obtained from the eigenvectors of
$\rho=L^{-1}y$.

\begin{figure}[ht]
    \centering
    \includegraphics[width=\linewidth]{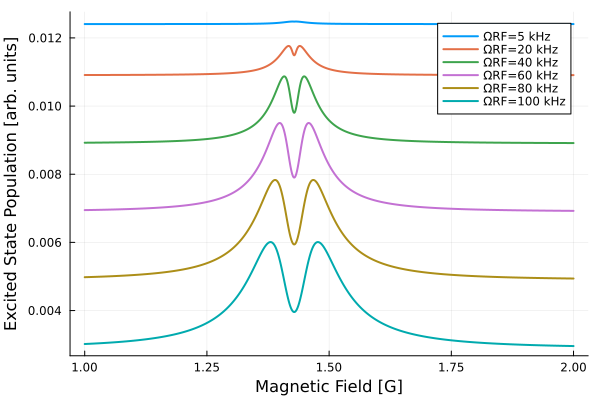}
    \caption{Optical-RF double resonances are calculated for various Rabi frequencies. $J=1 \longrightarrow J=0$ transition with $\mathbf{E} \parallel \mathbf{B}$.}
    \label{fig:calcs-double-res}
\end{figure}

\begin{figure}[ht]
    \centering
    \includegraphics[width=\linewidth]{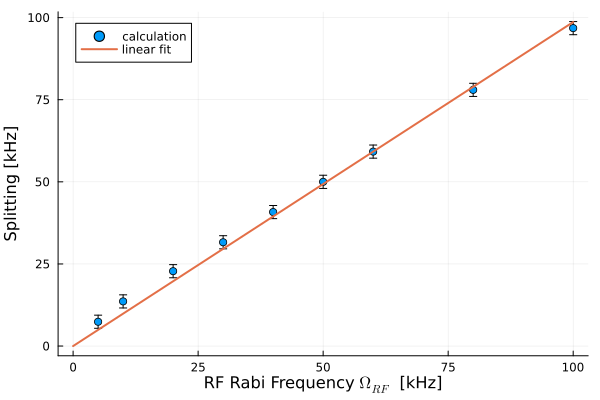}
    \caption{The calculated splitting of the peaks as a function of RF field Rabi frequency is shown. $J=1 \longrightarrow J=0$ transition with $\mathbf{E} \parallel \mathbf{B}$. 
    }
    \label{fig:calcs-AT-splitting-vs-Rabi}
\end{figure}

Figure~\ref{fig:calcs-double-res} shows the excited-state population versus the external magnetic field for a fictitious $J=1 \longrightarrow J=0$ transition for various RF Rabi frequencies over a range of 5 kHz to 100 kHz.   The excited-state population is identical to absorption in this case since each atom in the excited state is there because a photon was absorbed. Optical-RF double resonance peaks are observed at 1.45 kHz as expected.  AT splitting can be observed for all peaks, although at 5 kHz it is so small that peak positions cannot be determined accurately. Fig.~\ref{fig:calcs-AT-splitting-vs-Rabi} shows the splitting as a function of Rabi frequency. As can be seen, it is linear, just as in the experiments.

\section{Conclusion}
Autler-Townes splitting was observed in the optical-RF double resonance peaks when exciting the Cs D$_1$ transition with optical radiation whose polarization vector $\mathbf{E}$  was parallel as well as perpendicular to the external magnetic field. The RF magnetic field oscillated in the plane perpendicular to the optical field's electric polarization vector and in the case of $\mathbf{E} \perp \mathbf{B}$, moreover, was oriented in such as way as to have both $q=0$ and $q= \pm 1$ polarization components, which would allow two photons to induce a $\Delta m=1$ transition. Indeed, a peak was observed at a magnetic field value of $B=2 \cdot h \nu_{RF}/ (g_F \mu_B)$, which would correspond to a two-photon transition. Here $\nu_{RF}=\omega_{RF}/(2\pi)$. The Autler-Townes effect for a single-photon transition was characterized by a linear dependence of the splitting on the RF magnetic field amplitude, which is proportional to the Rabi frequency, although the constant of proportionality varies depending on the magnetic sublevels involved in the transition. In the case of $\mathbf{E} \perp \mathbf{B}$, the peaks at $B=2 \cdot h \nu_{RF}/ (g_F \mu_B)$ were split as well, and the magnitude of the splitting increased quadratically with the RF field magnetic field amplitude.  Furthermore, for strong RF Rabi frequencies, an additional peak at a magnetic field that corresponds to half the resonance frequency was observed, which would imply a single-photon transition between non-adjacent magnetic sublevels. A simple model for a $J=1 \longrightarrow J=0$ transition confirms the basic features of the linear optical-RF AT effect. Further work is needed to model the suspected two-photon AT effect as well as to explain the potential three-photon transition observed  at $B=3 \cdot h \nu_{RF}/ (g_F \mu_B)$ for high RF magnetic field values in the $\mathbf{E} \parallel {B}$ spectrum, and the peak at $B=\frac{1}{2} \cdot h \nu_{RF}/ (g_F \mu_B)$ in the $\mathbf{E} \perp {B}$ spectrum.

\section*{Acknowledgments}
This work was supported by Latvian Council of Science Project Nr. lzp-2020/1-0180, “Compact 3-D magnetometry in Cs atomic vapor at room temperature”. We thank Juris Birznieks and Ludvigs Mikelsons for helping to remeasure some plots.

\newpage
\bibliography{CsD1RF}
\bibliographystyle{apsrev}

\end{document}